\documentclass[%
 reprint,
 amsmath,amssymb,
 aps,
]{revtex4-2}

\usepackage{graphicx}% Include figure files
\usepackage{dcolumn}% Align table columns on decimal point
\usepackage{bm}% bold math
\usepackage{xcolor} % for \textcolor
\usepackage[mathlines]{lineno}% Enable numbering of text and display math
\usepackage{caption}
\usepackage{comment}

\begin{document}

\preprint{APS/123-QED}

\title{
\textbf{Fast Neural Noise Stabilizes Associative Memory against Slow Synaptic Noise}
}% Force line breaks with \\

\author{Tomoki Kurikawa}
 \affiliation{Department of Complex and Intelligent Systems, Future University Hakodate, 116-2 Kamedanakano-cho,
Hakodate, Hokkaido, 041-8655, Japan}%Lines break automatically or can be forced with \\
\email{kurikawa@fun.ac.jp}
\author{Kunihiko Kaneko}%
\affiliation{
 The Niels Bohr Institute, University of Copenhagen, Blegdamsvej 17, Copenhagen, 2100-DK, Denmark
 }


\date{\today}% It is always \today, today,

\begin{abstract}
Neural systems operate under multi-timescale stochastic dynamics, yet reliably maintain stable functions. We investigate how fast neural fluctuations and slow synaptic variability asymmetrically influence the structural robustness of associative memories. 
While synaptic noise degrades memory formation, neural noise stabilizes the network against structural perturbations.
For sequential learning of multiple memories, the average recall performance is maximized at an intermediate neural-noise amplitude, whereas increasing synaptic noise rapidly degrades memory performance.
Using an adiabatic mean-field framework, we derive an analytical stationary distribution of stability factors. Our theory reveals that fast neural fluctuations push the system away from marginal stability, serving as an indispensable prerequisite for systemic stability.
%\begin{description}
%use the optional argument of the \verb+\item+ command to give the category of %each item. 
%\end{description}
\end{abstract}

\maketitle

Neural systems operate under inherently stochastic, multi-timescale dynamics. Neural activity fluctuates on fast timescales (milliseconds) \cite{Churchland2011,Goris2014}, while synaptic strengths evolve over much slower periods (seconds to hours) \cite{Moreno-Bote2014,Faisal2008,Kasai2021}. 
Despite this pervasive fluctuation, biological networks reliably maintain stable cognitive functions\cite{Driscoll2017,Kasai2021}, such as associative memory. 
A central question in the physics of living systems is how learning remains robust in such noisy environments and whether different sources of noise play distinct functional roles.

Noise-induced robustness has been investigated in associative-memory models. 
Previous studies have examined noise in inputs or neural activity as a mechanism for enhancing the robustness of associative and working memories \cite{Gardner1989,Benedetti2024,Wei2014,Karbasi2014,Zhang2021,Rungratsameetaweemana2025}, as well as synaptic or weight noise in relation to memory reliability and generalization \cite{Jim1996,Zhang2021,Ho2009}. 
These noise effects are also related to noise-induced regularization in machine learning \cite{Bishop1995,Lim2021}. 
Despite this progress, however, the respective roles of fast neural fluctuations and slow synaptic variability in memory stabilization remain unclear. 
Here, we disentangle these contributions and show that neural and synaptic noise have fundamentally different effects on the stabilization of learned input/output associations.

In this study, we show that neural and synaptic noise exert an asymmetric influence on the structural robustness of associative memories. 
While synaptic noise primarily degrades the quality of memory formation, neural noise during training acts as a crucial stabilizer. 
To provide a theoretical foundation for this asymmetry, we develop an adiabatic mean-field (AMF) framework for the neural-synaptic coupling dynamics\cite{Clark2024,Siri2008,Aoki2009}. 
By treating neural activity as a fast variable relative to slow synaptic evolution\cite{Gardiner2010,Haken1978,Risken1996,Kaneko1981} with mean-field method\cite{Sompolinsky1988,Kadmon2015}, we derive an analytical stationary distribution of the stability factors. 
Our theory reveals that neural noise effectively  "pushes" the system away from marginal stability, while the synaptic noise works as restoration force.  
This work clarifies how the disparate timescales of biological noise cooperatively shape the connectivity landscape to ensure functional robustness.

We study a recurrent neural network (RNN) of $N$ continuous-valued units $x_i(t)$ with a synaptic matrix $J_{ij}$ designed to memorize input/output (I/O) patterns. 
Self-connections are removed by enforcing $J_{ii}=0$. 
The neural state evolves according to the standard rate coding model:
\begin{equation}
\frac{dx_i(t)}{dt} = -x_i(t) + \tanh \left( \beta \left( \sum_{j \neq i} J_{ij} x_j(t) + \eta_i \right) \right) + \zeta^n_i(t).
\label{eq:neural}
\end{equation}
White Gaussian neural noise $\zeta^n_i(t)$ is included, satisfying $\langle \zeta^n_i(t) \zeta^n_j(t') \rangle = 2 D_n \delta_{ij} \delta(t-t')$, where $D_n$ represents the neural noise intensity. 
Input and target output patterns, $\boldsymbol{\eta}$ and $\boldsymbol{\xi}$, are $N$-dimensional vectors with values randomly chosen from $\{-1,1\}$. In this study, we set $N=1000$ and $\beta=10$.

To learn the pattern, each synapse evolves according to a homeostatic error-corrective Hebbian rule \cite{Kurikawa2020} with a characteristic timescale $\tau_J$:
\begin{equation}
\tau_J \dot{J}_{ij} = \frac{1}{N} (\xi_i - x_i) x_j + \alpha (1 - R_i) J_{ij} + \zeta^w_{ij}(t),
\label{eq:synaptic}
\end{equation}
where $R_i \equiv \sum_{k} J_{ik}^2$ is the squared row-norm of synapses from neuron $i$. 
The first term - error-corrective Hebbian term- in Eq.~(\ref{eq:synaptic}) drives $x_i$ toward the target $\xi_i$ via a process akin to Perceptron learning, while the second term—the homeostatic term—stabilizes total synaptic strength around unity \cite{Miller1994,Hyva1998,Zenke2017b}. 
This rule is a first-order model of pre- and post-synaptic activity, representing the minimum complexity required to produce robust I/O maps \cite{Kurikawa2013, Kurikawa2016}. 
The stabilization amplitude is controlled by $\alpha=0.001$, with $\tau_J=10$. 
Synaptic noise $\zeta^w_{ij}(t)$ is white Gaussian with intensity $D_s$, satisfying $\langle \zeta^w_{ij}(t) \zeta^w_{i'j'}(t') \rangle = 2 D_s N^{-1} \delta_{ii'}\delta_{jj'} \delta(t-t')$.
$J_{ij}$ is initialized according to i.i.d of $\mathcal{N}(0,N^{-1})$.

During learning, the dynamics run in the presence of $\boldsymbol{\eta}$ until the network reaches a stationary state, quantified by the overlap $m \equiv [\xi_i x_i]_i$, where $[\cdot]_i$ denotes the population average. 
As shown in Fig.~1, neurons exhibit irregular fluctuations during the transient phase. Despite this variability, the overlap $m$ increases almost monotonically toward unity as learning progresses.

\begin{figure}[t]
\centering
\includegraphics[width=0.40\textwidth]{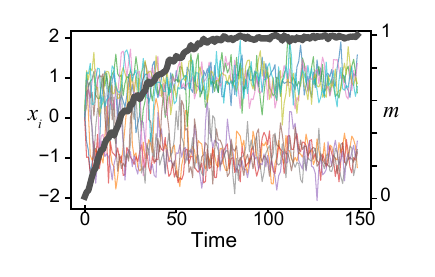}
\caption{Dynamics of 10 neural activities $x_i$ out of $N=1000$ (colored lines) and overlap $m$ (black line) during learning for $(D_s,D_n)=(0,0.1)$.}
\label{fig:fig1}
\end{figure}

We first investigate how stochasticity affects learning by measuring the temporally averaged overlap $m$. 
The neural dynamics were simulated for \(0 \leq t \leq 32000\), and the overlap was averaged over $12500\le t \le 32000$ after a sufficiently long buffer period following convergence of the neural state toward the target pattern.

When both noises are very weak  ($D_s = D_n = 10^{-2}$), the overlap reaches unity, as shown in Fig. 2A.
In this study, we set the minimum noise at $10^{-2}$ because the synaptic dynamics without noise do not converge into the stationary state (see Appendix 1).
While increasing noise reduces the asymptotic overlap, performance remains remarkably high; even at $D_s = D_n = 1$, the overlap exceeds $0.9$, indicating that the learning process is robust to high levels of stochasticity.
Note the asymmetric effect of synaptic and neural noises in the overlap $m$; $D_s$ strongly decreases $m$ compared to $D_n$.

A central claim of this study is that noise during training fundamentally enhances the structural robustness of the resulting fixed point $\boldsymbol{\xi}$. 
To isolate the intrinsic dynamical properties of the learned connectivity, we examine the recall process with the synaptic matrix frozen and the noise sources removed. 
We quantify structural robustness by introducing a perturbation $J_{ij} \to J_{ij} + \Delta J_{ij}$, where $\Delta J_{ij} \sim \mathcal{N}(0, p^2/N)$, and measuring the steady-state recall overlap as a function of the perturbation amplitude $p$.
During the recall process, the neural state converges to the target quickly and we numerically simulated the dynamics for $0\le t \le 50$.
The overlap is averaged over $48\le t  \le 50$. 

The impact of neural noise $D_n$ on stability (at $D_s = 10^{-2}$) is shown in Fig.2B. 
For networks trained without noise ($D_n = 10^{-2}$), the overlap drops precipitously with $p$. 
In this regime, even small perturbations render the fixed point $\boldsymbol{\xi}$ unstable, causing the system to collapse into limit cycles or chaotic attractors. 
Conversely, training with larger $D_n$ dramatically enhances the local stability of the fixed point, allowing the network to maintain high recall performance under significant structural perturbation.

We define the critical perturbation amplitude $p_c$ as the threshold where recall overlap falls below $0.95$. 
Figure 2C presents $p_c$ as functions of $D_s$ and $D_n$. 
The results reveal a striking asymmetry: while neural noise $D_n$ confers robustness independent of $D_s$, synaptic noise $D_s$ alone fails to improve stability. 
Actually, even with neural noise present (e.g., $D_n=1$ in Fig. 2C), $D_s$ contributes quite little to stability.

\begin{figure*}[t]
\centering
\includegraphics[width=0.9\textwidth]{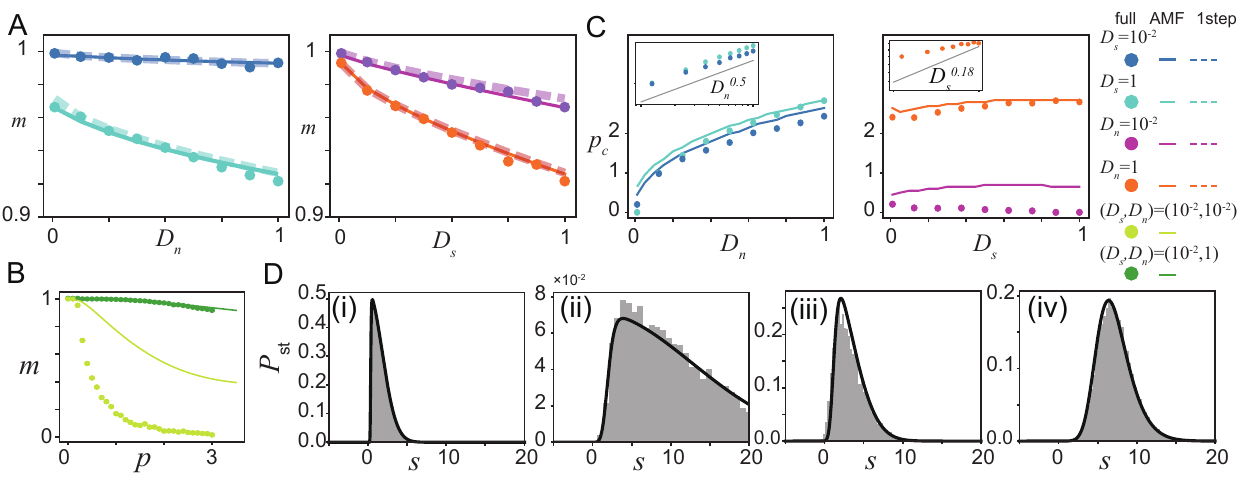}
\caption{A and C: $m$ and $p_c$ as functions of  $D_n$ (left) and $D_s$ (right), respectively. 
Insets of C show the corresponding data on a log-log scale, highlighting the scaling relations. 
B: Recall overlap $m$ vs. perturbation $p$. 
The color code in panels A-C is shown on the right: full, AMF, and 1 step represent full simulation, adiabatic mean-field approximation, and 1-step approximation, respectively.
D: Distribution of stability factors $s_i$ after training for $(D_s,D_n)=(10^{-2},10^{-2}),(10^{-2},1),(1,10^{-2})$ and $(1,1)$ in (i)-(iv), comparing AMF theory (black line) to simulation (gray). 
All full-simulation results were averaged over 10 independently initialized networks, each trained on a distinct input/output map.
}
\label{fig:fig2}
\end{figure*}

To elucidate the mechanism underlying this robustness, we analyze the stability factor $s_i$, defined for each neuron $i$ as:
\begin{equation}
s_i \equiv \xi_i \left( \sum_j J_{ij} \xi_j + \eta_i \right).
\label{eq:stability_factor}
\end{equation}
In the zero-temperature limit ($\beta \to \infty$), the configuration $\boldsymbol{\xi}$ is a stable fixed point if and only if $s_i > 0$ for all $i$. 
$s_i$ quantifies the alignment between the state of neuron $i$ and its local field, $h_i = \sum_j J_{ij} \xi_j + \eta_i$.
Its magnitude serves as a direct measure of the local stability margin \cite{Amit1985,Gardner1988, Kurikawa2026}; 
larger values of $s_i$ indicate a more robust fixed point, as a substantial margin ensures that the sign of the local field remains invariant under perturbations to the connectivity $J$ or the external field $\boldsymbol{\eta}$.

The distribution of $s_i$ evolves significantly during learning and is shaped differentially depending on $D_s$ and $D_n$. 
In the very weak noise case ($D_n = D_s = 10^{-2}$), the distribution shaped through learning is sharply peaked just above the threshold $s = 0$ (Fig.~2D(i)). 
Such a configuration is extremely fragile:  even small perturbations destabilize the target.
A similar concentration near $s=0$ is observed for synaptic noise alone ($D_s=1, D_n=10^{-2}$), consistent with the weak robustness observed above.  
In stark contrast, neural noise during learning reshapes the distribution, shifting the probability mass toward larger $s$ and suppressing the density near the marginal boundary $s = 0$ (Fig.~2D(ii, iv)). This creates a "stability margin," ensuring individual neurons possess a buffer to withstand connectivity fluctuations.
These results demonstrate that the noise-induced robustness observed in Fig.~2C is a direct consequence of this learned stability profile, which effectively "deepens" the potential well of the target fixed point.

To provide a theoretical foundation for the observed robustness, we develop an adiabatic mean-field (AMF) approximation describing the slow evolution of the stability factors $s_i$ and the row-norms of weights, $R_i$. 
By employing an adiabatic approximation on the stochastic system \cite{Kaneko1981,Risken1996}, we treat the neural activity $x_i$ as a fast variable that reaches a local equilibrium relative to the slow synaptic changes $J_{ij}$, combined with mean-field approach for $\boldsymbol{x}$  in the thermodynamic limit ($N \to \infty$). 
We rescale time such that neural dynamics evolve on a timescale $1/\tau_J$, while synaptic dynamics evolve on a unit timescale.

The evolutions of $s$ and $R$ are governed by the following reduced equations (see Appendix 1):
\begin{align}
    \frac{ds}{dt} &= \left(1 - \langle\tanh(\beta u)\rangle_{u|s}\right) m + \alpha (1 - R)s + \zeta^s, \label{eq:s_dyn_mf} \\
    \frac{dR}{dt} &= 2\alpha (1 - R) R + \frac{2D_s}{\tau_J}, \label{eq:R_dyn_mf}
\end{align}
where $\zeta_i^s$ is  a Gaussian white noise term with $\langle \zeta^s_i(t) \zeta^s_j(t') \rangle = 2D' \delta_{ij}\delta(t-t')$, where
\begin{eqnarray}
    D'=(D_s+m^2 D_n)\tau_J^{-1}.
\end{eqnarray}
$<\tanh(\beta u)>_{u|s}=\int \tanh(\beta u) P(u|s) du$ with 
\begin{equation}
    P(u|s)=\mathcal{N}( s, \Sigma^2 = R(D_n + 2(1-m))). \label{eq:u_distr}
\end{equation}
This distribution of the local field \(u\) is centered at the stability factor \(s\), with a variance determined by the neural noise \(D_n\) and by the deviation of the current activity from the target pattern. 
Averaging the activation function over this distribution captures the effective broadening induced by neural fluctuations, which shifts the stability factors away from the marginal boundary. 
Details of this derivation are given in Appendix~2.

Using the reduced dynamics, we obtain the analytical stationary distribution $P_{\text{st}}(s)$. 
In this limit, $R_i$ converges to a deterministic value $R_{\text{st}}$:
\begin{equation}
R_{\text{st}} = \frac{1}{2} + \sqrt{\frac{1}{4} + \frac{D_s}{\alpha \tau_J}}.
\label{eq:Rst}
\end{equation}

By substituting $R_{\text{st}}$ into Eq.~(\ref{eq:s_dyn_mf}) and averaging the activation function over the local field distribution $u_i$, we derive the Fokker-Planck equation for the probability density $P(s,t)$:
\begin{equation}
    \frac{\partial P(s,t)}{\partial t}= -\frac{\partial}{\partial s} \left[ F(s) P(s,t) \right] + D' \frac{\partial^2 P(s,t)}{\partial s^2}.
\label{eq:fpe}
\end{equation}
The drift force $F(s) = m(1 - \langle \tanh(\beta u) \rangle_{u|s}) - \gamma s$ balances the Hebbian drive with the homeostatic restoration, where 
\begin{equation}
    \gamma = \alpha(R_{\text{st}} - 1).\label{eq:gamma}
\end{equation}

The stationary overlap $m_{\text{st}}$ is determined via a self-consistent iterative scheme\cite{Gummel1964,Kerker1981}. 
Given a value of $m$, we solve for $P_{\text{st}}(s|m)$ and update $m_{\text{new}}= \int \langle \tanh(\beta u) \rangle_{u|s} P_{\text{st}}(s|m) ds$ \footnote{Considering $m=\Sigma_i \xi_i x_i/N $ and the fixed point condition $x=\tanh(\beta Jx + \eta)$, $m=\Sigma_i \tanh(\beta u_i)/N \rightarrow \int \tanh(\beta u) P(u|s)P(s) du ds$}. 
Here, the initial value of $m$ is set at $1$.
Solving Eq.~(\ref{eq:fpe}) for the steady state, we obtain the Boltzmann-like form:
\begin{equation}
P_{\text{st}}(s|m) = \frac{1}{\mathcal{Z}} \exp\left( -\frac{V(s|m)}{D'} \right),
\label{eq:stationary_dist}
\end{equation}
where the effective potential $V(s|m)$ is given by:
\begin{multline}
V(s|m) =  \\ 
\frac{\gamma}{2} s^2 - m \left[ s - \left(s \cdot \text{erf}\left(\frac{s}{\sqrt{2}\Sigma}\right) + \sqrt{\frac{2}{\pi}}\Sigma \exp\left(-\frac{s^2}{2\Sigma^2}\right)\right) \right].
\label{eq:potential}
\end{multline}
The first term in $V(s|m)$ represents a parabolic restoring force originating from synaptic regularization.
The second term provides a ``pushing force'' driving $s$ toward positive values. 
Crucially, the presence of $\Sigma$ (linked to $D_n$) in the error function term smoothens the potential landscape, preventing the distribution from becoming a Dirac delta pinned at the zero-boundary.

Further, we evaluate the structural robustness $p_c$ within the AMF framework. 
A synaptic perturbation $J_{ij} \to J_{ij} + \Delta J_{ij}$ with $\Delta J_{ij} \sim \mathcal{N}(0, p^2/N)$ induces a Gaussian broadening of the stability factor distribution.
By convolving $P_{\text{st}}(s)$ with a Gaussian kernel of width $p$, we obtain the perturbed distribution and the overlap $m(p)$ with this distribution. 
As shown in Fig.~2C, the AMF theory faithfully reproduces the $p_c$ observed in simulations, where robustness is generally enhanced as both $D_s$ and $D_n$ increase.
However, a fundamental asymmetry remains: at $D_s \approx 0$, increasing $D_n$ significantly strengthens the robustness, whereas at $D_n \approx 0$, increasing $D_s$ yields negligible improvement.

While the AMF theory provides a comprehensive description of the stationary distribution $P_{\text{st}}(s)$, the resulting expressions for the overlap $m$ and the robustness $p_c$ are mathematically complex.
To extract the essence of how neural and synaptic noise govern these quantities, we derive simplified scaling relations. 

First, we consider the stationary overlap $m$. By evaluating the self-consistent iteration only once, we obtain an approximate form $m^{(1)}$:
\begin{equation}
m^{(1)} = 1 - \gamma \left( \sqrt{\frac{\pi}{2}}\Sigma + \sqrt{\frac{2D'}{\pi\gamma}} \right),
\label{eq:m1}
\end{equation}
where $\Sigma^2, D'$, and $\gamma$ are approximated as $D_n\sqrt{D_s /\alpha \tau_J}$, $(D_s+D_n)/\tau_J$, and $\sqrt{\alpha D_s/\tau_J}$, respectively (we set $m=1$) as in Eq. \eqref{eq:control_params}. 
This simple formulation of $m$ is remarkably applicable to the full simulation results, as shown in Fig.~2A.
According to Eq.~(\ref{eq:m1}), the overlap deficit scales as $1-m \propto D_n^{0.38}$ for constant $D_s=1$, while it follows $1-m \propto D_s^{0.63}$ for constant $D_n=1$ (see Appendix 3). 
This theoretical evaluation clarifies that \(D_s\) decreases \(m\) more strongly than \(D_n\)
as shown in the full simulation.

Additionally, we characterize the structural robustness $p_c$ by identifying the ``rising point'' of the distribution $P_{\text{st}}$ from zero. 
By calculating the value $s'$ that satisfies $V(s') \sim D'$ in Eq.~(\ref{eq:stationary_dist}), we derive a simplified proportional form for $p_c$:
\begin{equation}
    p_c \propto \sqrt{\frac{\pi}{2}}\Sigma - \sqrt{\sqrt{2\pi} D' \Sigma}. \label{eq:p_c}
\end{equation}
Although we do not evaluate the absolute value of $p_c$ directly, this proportionality captures the essential dependence on $D_s$ and $D_n$ observed in Fig.~2C. 
Specifically, we find that $p_c \propto D_n^{0.5}$ when $D_s$ is fixed. In the absence of neural noise ($D_n = 0$), $p_c$ remains near zero regardless of $D_s$, whereas for finite neural noise (e.g., $D_n=1$),
\(p_c\) scales weakly as \(D_s^{0.18}\), as obtained numerically.
This simple formulation from AMF theory clarifies the asymmetric impact of neural and synaptic noise on the robustness of the target state. 
Our theory thus provides a quantitative framework for how neural and synaptic noise cooperatively reshape the connectivity landscape to enhance functional robustness.

Finally, we examined how neural and synaptic noise affect the stability of multiple memories. 
We trained the network sequentially on random binary I/O pairs 
\((\eta^\alpha,\xi^\alpha)\) \((\alpha=0,\ldots,N_{\rm pat}-1)\), using the same learning protocol as above. 
Once the overlap with the current target reached \(m=0.99\), the next pair was presented and learned. 
In these simulations, \(N=200\) and \(N_{\rm pat}=20\). 
After all patterns had been learned, we measured the recall overlap 
\(m^\alpha\equiv [\xi_i^\alpha x_i]_i\) for each target pattern \(\xi^\alpha\) at different values of \(D_n\) with \(D_s=0.01\) (Fig.~3A). 
The recall dynamics were simulated for $0\le t \le 100$ for each input and the corresponding overlap was measured over $50\le t \le 100$. 
Because learning is sequential, later updates interfere with earlier memories; thus, early learned patterns show reduced recall for all values of \(D_n\) and \(D_s\) tested.
This is true even in the noiseless ($D_n=D_s=0$) case.

The extent of this degradation depends strongly on the noise amplitudes, because the stability margin of each memory is controlled by \(D_n\) and \(D_s\). 
For \(D_s=0.01\), the total memory performance, or capacity, 
\(C\equiv N_{\rm pat}^{-1}\sum_{\alpha}m^\alpha\), is maximal around \(D_n\simeq0.25\) (Fig.~3A,B). 
Increasing \(D_n\) stabilizes the most recently learned pattern, consistent with Fig.~2C. 
However, forming a larger stability margin for the current pattern requires a larger rearrangement of the synaptic matrix, which more strongly erodes the stability of previously learned patterns. 
This trade-off yields an optimal neural-noise amplitude for the collective performance of multiple memories.
Actually, for $D_n=0.25$ and $0.5$, the overlaps with recently learned patterns are much higher. 
As $D_n$ is increased beyond $0.25$, however, the overlaps with early learned patterns are further reduced (Fig. 3A), resulting in an optimum of the capacity near $D_n=0.25$ (Fig. 3B).

For \(D_s=1\), synaptic noise induces much larger synaptic fluctuations during learning. 
Interference with previously learned memories is therefore much stronger than for \(D_s=0.01\), and the capacity remains low over the entire range of \(D_n\), with no clear optimum. 
Consistently, increasing \(D_s\) at fixed \(D_n\) rapidly decreases the capacity (Fig.~3C), reflecting stronger synaptic-noise-induced interference with previously learned memories.

\begin{figure}
    \centering
    \includegraphics[width=1\linewidth]{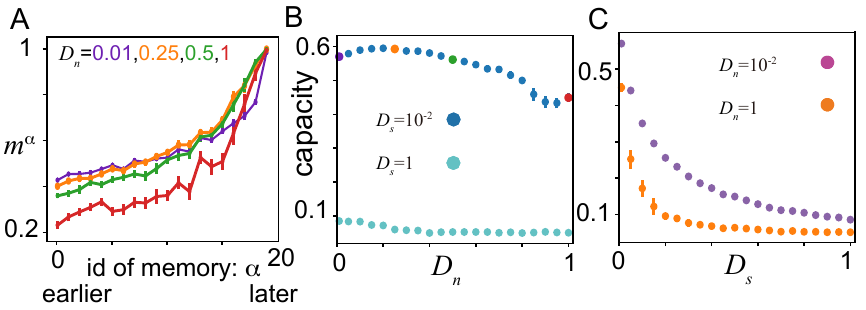}
    \caption{A: Overlap with $\boldsymbol{\xi^\alpha}$, $m^{\alpha}$ is plotted for different $D_n$ with $D_s=0.01$. 
    B and C: The capacity for different $D_n$ and $D_s$ is plotted, respectively. The capacity is defined as $\Sigma_\alpha m^{\alpha}/N_{\text{pat}}$. All results were averaged over 150 trials: five trials from different initial states for each of 30independently initialized networks, with each network trained on distinct input/output maps. The error bars indicate the standard error of the mean across 30 networks.} 
    \label{fig:fig3}
\end{figure}

In this study, we have demonstrated that neural and synaptic noise play asymmetric roles in the formation and stabilization of associative memories. 
While synaptic noise \(D_s\) primarily degrades the memory overlap \(m\), neural noise \(D_n\) during learning makes the target state robust to subsequent structural perturbations. 
This result indicates that a neural system is inherently unstable in the presence of synaptic fluctuations alone; fast neural fluctuations are an indispensable prerequisite for systemic stability.
This asymmetry is captured by our adiabatic mean-field (AMF) framework, which shows that neural fluctuations push the stability factors away from marginality through the neural-noise-dependent broadening \(\Sigma\) of the effective activation function. 
In contrast,  synaptic noise exerts competing effects on stability: it also broadens $\Sigma$ (albeit more weakly than neural noise) while simultaneously acting as a restoring force through $\gamma$, which stems from synaptic scaling.
While previous studies have shown that finite stability margins and learning from noisy examples can enlarge attractor basins \cite{Gardner1988,Gardner1988b,Gardner1989}, they have not clarified the respective contributions of fast neural fluctuations and slow synaptic variability. 
Although multiple-timescale networks have been studied previously \cite{Zhang2021,Clark2024,Siri2008,Magnasco2009,Teramae2022}, our AMF framework analytically separates these two sources of stochasticity, which demonstrates that they have distinct effects on the learned stability distribution. 
Because neural fluctuations have also been linked to responses to external inputs \cite{Cessac2019} and to learning speed \cite{Kurikawa2025,Sadtler2014}, our results may provide a general framework to elucidate  the distinct roles of neural and synaptic noise for robustness, responsiveness, and learning speed.

Such noise-induced structural robustness finds a compelling parallel in the study of biological evolution\cite{Kaneko2025}. 
In gene expression models, phenotypic noise during development has been shown to facilitate the evolution of 'fitted' states that are intrinsically resilient to mutations that alter network structure \cite{Siegal2002,Ciliberti2007,Kaneko2007}. 
This mirrors our results, suggesting a shared physical principle where 'fast' fluctuations buffer 'slow' adaptive changes across different biological scales. 
Formally unifying these multi-timescale processes—potentially via adaptive dynamical mean-field theory \cite{Pham2024}—offers a promising path toward a universal theory of robustness in adaptive systems.

The AMF method presented here is not limited to the specific model studied but is potentially applicable to a broad class of multiple-timescale stochastic systems. 
The separation of timescales—where synapses evolve slowly relative to the fast equilibrium of neural states—allows for the reduction of high-dimensional stochastic dynamics into a tractable Fokker-Planck description of the stability factor distribution. 
This suggests that the "noise-induced robustness" observed here may be a universal feature of learning in multi-timescale stochastic networks.

We further analyzed sequential learning of multiple patterns.
For weak synaptic noise, the total recall performance was maximized at an intermediate neural-noise amplitude.
This optimum reflects a trade-off: neural noise stabilizes the currently learned memory, as predicted by the single-memory AMF analysis, but excessive stabilization requires larger synaptic rearrangements and thereby accelerates the decay of previously learned memories.
For stronger synaptic noise, the overlaps of earlier memories decayed rapidly and no clear optimum was observed.
The present multiple-memory results are numerical. Extending the AMF framework to sequential learning would require characterizing how each subsequent learning episode shifts and broadens the stability distribution of previously stored memories. Once these episode-to-episode transition statistics are specified, the AMF description can in principle be propagated across successive learning steps. Deriving these transition statistics self-consistently, however, would require an extended mean-field theory that explicitly incorporates interference among sequentially learned memories, which is beyond the scope of the present study.

\begin{acknowledgments}
The authors thank Natsuhiko Yoshinaga for helpful discussions and valuable comments on this work.
This work is supported by JSPS KAKENHI (Nos 26K14982, 25H02622 and 25K02550 for T.K.), NNF21OC0065542 (K.K.), and Special Research Expenses in Future University Hakodate
(T.K.).
\end{acknowledgments}

\bibliography{25th_rev}% Produces the bibliography via BibTeX.

\clearpage
\newpage
\section*{End matter}

\subsection*{Appendix 1: Derivation of Adiabatic Mean Field (AMF) Dynamics and Stationary Distributions}

In this section, we derive the dynamics of the squared row-norm $R_i = \sum_j J_{ij}^2$ and the stability factor $s_i = \xi_i \left( \sum_j J_{ij} \xi_j + \eta_i \right)$ in mean field approximation and adiabatic approximation.
Here, we rescale the timescale of $J_{ij}$ to unity (namely, the temporal and stochastic increments in Eq. \eqref{eq:neural} and \eqref{eq:synaptic} are rescaled as $dt/\tau_J \to dt$ and $dW/\sqrt{\tau_J} \to dW'$ ) and consequently
$\frac{dx}{dt} = \tau_J (\tanh(\beta (Jx+\eta))-x) +\sqrt{\tau_J}\zeta^n $ and $\frac{dJ}{dt}=\frac{1}{N}(\xi-x)x + \alpha(1-R)J+\sqrt{\tau_J}^{-1}\zeta^w$.

First, we consider $R_i$ according to Eq. \eqref{eq:synaptic}.
Applying Itô's lemma to the synaptic update rule, the differential $dR_i$ is expanded as $dR_i = 2 \sum_j J_{ij} dJ_{ij} + \sum_j (dJ_{ij})^2$.
The second-order Itô shift yields a strictly positive constant drift, $\sum_j (dJ_{ij})^2 = 2D_s \tau_J^{-1} dt$.
In the thermodynamic limit ($N \to \infty$), the variance of the effective noise on $R_i$ ($=\Sigma_j J_{ij}\zeta^{w}_j$) scales as $\mathcal{O}(N^{-1})$ and vanishes, ensuring that the macroscopic dynamics are purely deterministic.
Here, we use $J_{ij} = \mathcal{O}(N^{-1/2})$ uncorrelated with $\zeta^{w}$.
Furthermore, the cross-terms originating from the Hebbian update, $N^{-1} \sum_j (\xi_i - x_i) x_j J_{ij}$, scale as $\mathcal{O}(N^{-1})$ and also vanish given that we roughly evaluate the correlation between $J_{ij}$ and $x_i, x_j$ is quite small. 
Dropping the index $i$, the dynamics reduce to:
\begin{equation*}
\frac{dR}{dt} = 2\alpha R(1 - R) + \frac{2D_s}{\tau_J}
\end{equation*}
The stationary state $R_{\text{st}}$ is obtained by  $dR/dt = 0$.
Solving the resulting quadratic equation and discarding the negative root yields the fixed point:

\begin{equation*}
R_{\text{st}} =\frac{1}{2}\left(1 + \sqrt{1 + \frac{4D_s}{\alpha\tau_J}}\right) \sim
\begin{cases}
\sqrt{\frac{D_s}{\alpha \tau_J}}, & D_s \gg 10^{-2}, \\
1, & D_s \ll 10^{-2}.
\end{cases}
\end{equation*}

because $\alpha \tau_J $  $=10^{-2}$.

Next, we derive the adiabatic dynamics of $s_i$, Eq.\eqref{eq:s_dyn_mf}.
Based on Eq.\eqref{eq:synaptic} , the dynamics of $s_i$ are:
\begin{eqnarray}
    \frac{ds_i}{dt} &=& \sum_j  \frac{1}{N} (1 -\xi_i x_i) x_j\xi_j - \gamma (s_i-\xi_i\eta_i)  + \nu, \label{eq:s_dyn}
\end{eqnarray}
where $\nu$ is a slow noise as $\nu=\frac{1}{\sqrt{\tau_J}}\xi_i \Sigma_j \xi_j \zeta^w_{ij}$  and $\gamma=\alpha(R -1)$ as in Eq. \ref{eq:gamma}. 
Assuming a separation of timescales between $x$ and $J$, we derive $ \frac{1}{N} (1 -\xi_i x_i) x_j\xi_j = (1-\langle\tanh(\beta u)\rangle_{u|s_i})m + (m/\sqrt{\tau_J})\zeta^n $ by using the adiabatic and mean-field approximation  (see Appendix 2), where  $<\tanh(\beta u)>_{u|s_i}=\int\tanh(\beta u)P(u|s_i)du$ with $P(u|s_i) = \mathcal{N}(s_i, \Sigma^2=R(2(1-m)+D_n))$ .

The  noise $\zeta^s_i(t) = \sqrt{\tau_J}^{-1}(m\zeta^n_i(t)+\xi_i \sum_j \xi_j \zeta^w_{ij}(t))$ is $<(\zeta^s_i)^2 >=2D'=2(D_s+m^2D_n)/\tau_J$.
In the following, we drop the index $i$, while the effect of the binary random variable $\xi_i \eta_i$ is canceled out by averaging over the index $i$ in the mean field approximation.
For large $\beta$ (e.g., $\beta = 10$), $\tanh(\beta u)$ is well approximated by a step function, and the expectation over the Gaussian distribution resolves to the error function:
\begin{equation}
\langle \tanh(\beta u) \rangle_{u|s} \approx \text{erf}\left(\frac{s}{\sqrt{2}\Sigma}\right). \label{eq:approx_tanh}
\end{equation}

Assuming the norm has reached its stationary state ($R = R_{\text{st}}$), the Langevin equation governing  $s$ becomes:
\begin{eqnarray*}
\frac{ds_i}{dt} 
&=& m\left(1 - \text{erf}\left(\frac{s}{\sqrt{2}\Sigma}\right)\right) -\gamma s + \zeta^s(t).
\end{eqnarray*}

Then, we obtain the stationary distribution of $s$.
The Langevin dynamics map directly to a Fokker–Planck equation for the probability density function $P(s,t)$. The stationary distribution $P(s)$ characterizing the fixed points of the network takes the Boltzmann-like form:
\begin{equation}
P(s) = \frac{1}{Z}\exp\left(-\frac{V(s)}{D'}\right) ,
\end{equation}
where $Z$ is the partition function.
The effective potential $V(s)$ is obtained by integrating the deterministic drift terms.
The integration of the error function yields an analytic closed form, providing the full effective potential:
\begin{widetext}
\begin{equation}
V(s) =\underbrace{ -ms + m\left[s\cdot \text{erf}\left(\frac{s}{\sqrt{2}\Sigma}\right) + \sqrt{\frac{2\Sigma^2}{\pi}}\exp\left(-\frac{s^2}{2\Sigma^2}\right)\right]}_{\text{from Hebbian part}} + \underbrace{\frac{\gamma}{2}s^2}_{\text{from synaptic scaling}}.
\end{equation}
\end{widetext}

\begin{figure}
    \centering
    \includegraphics[width=0.9\linewidth]{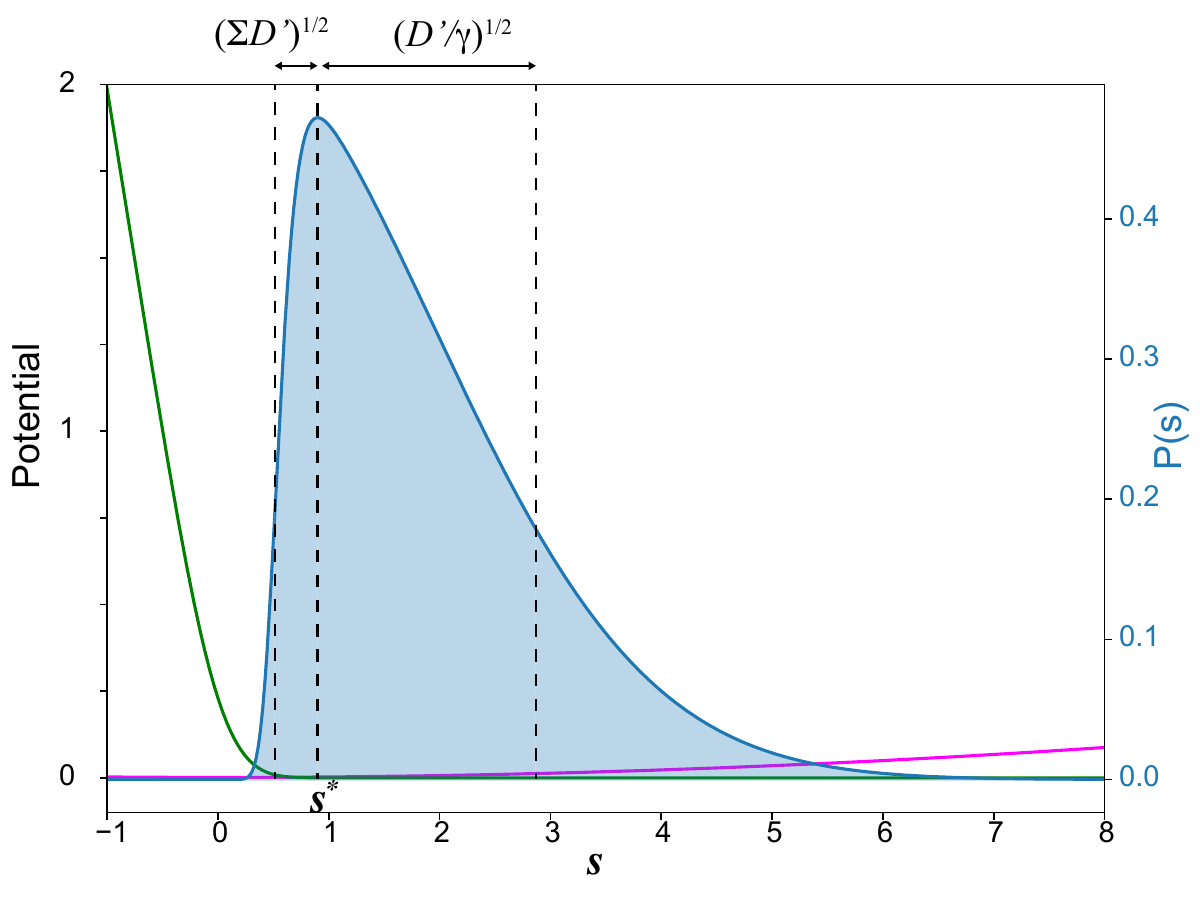}
    \caption{The potential and stationary distribution of $s$. Green and magenta lines represent the Hebbian and synaptic scaling parts, respectively. Shaded area indicates the stationary distribution.}
    \label{fig:s_distr}
\end{figure}

$P(s)$ is composed of two parts: for $s<s^*$, the Hebbian part is dominant, while for $s^*<s$, the synaptic scaling part is dominant, where $s^*$ is the local maximum of $P(s)$.
The width of the first part is evaluated as $(\Sigma D')^{1/2}$, while that of the second part is $(D'/\gamma)^{1/2}$ as shown in Fig. \ref{fig:s_distr}.
$\Sigma^2$, $D'$, and $\gamma$ depend on $D_s$ and $D_n$ as
\begin{equation}
\begin{aligned}
\Sigma^2 &\sim 
\begin{cases}
D_n \sqrt{D_s/(\alpha \tau_J)}, & D_s\ge10^{-2} \\
D_n, & D_s\le 10^{-2}  
\end{cases}\\  
D'&= (D_s+D_n)/\tau_J, \\
\gamma &\sim \sqrt{\alpha D_s/\tau_J}.
\end{aligned}
\label{eq:control_params}
\end{equation}

Given that when $D_s$ and $D_n$ increase from $0$ to $1$, $\Sigma^2$, $D'$, and $\gamma$ vary from $0$ to $10$, $0$ to $10^{-1}$, and $0$ to $10^{-2}$, respectively,
the first part becomes steep like a wall especially for smaller noise, while the second part resembles a half-Gaussian distribution with $\mathcal{O}(1)$ width.
Further, by expanding the derivative of $V(s)$ to first order in $s/\Sigma$, $V'(s) \sim -m+\sqrt{2/\pi}(ms/\Sigma) + \gamma s$,
we derive $s^* = m/(\gamma+m\sqrt{2/\pi}\Sigma^{-1}) \sim \sqrt{\pi/2}\Sigma$ due to $\gamma \ll \Sigma^{-1}$ and $m=1$.
When confirming stationary distribution derived from AMF dynamics with that from full simulation, we set the minimum noise strength as $10^{-2}$, because  the potential $\gamma s^2/2$  gets flat and the time to converge to the stationary distribution gets much longer when $D_s \to 0$.

\subsection*{Appendix 2: Adiabatic and Mean Field Approximation}

Here, we evaluate the fast term $\Sigma_j\frac{1}{N} (1 -\xi_i x_i) x_j\xi_j $ in the slow dynamics of $s$  in Eq.\eqref{eq:s_dyn}.
In the slow timescale, the dynamics of $x$ following Eq. \eqref{eq:neural} shape a stationary Gaussian distribution around $x^0$, $\mathcal{N}(x^0,D_n)$ , denoted as $\bar{x}$ where $x^0$ is a fixed point without noise for a given configuration of $J$, $x^0=\tanh(\beta (Jx^0+\eta))$.
In the fast scale, $x$ fluctuates around $\bar{x}$ driven by the fast neural noise $\zeta^n$.

Based on the adiabatic elimination\cite{Kaneko1981,Risken1996}, the fast variable $x_i$ is replaced by $\bar{x}+\sqrt{\tau_J}^{-1}\zeta^n$.
Considering $\sum_i \xi_i (\bar{x}+\sqrt{\tau_J}^{-1}\zeta^n) /N \sim \sum_i \xi_i x^0_i/N =m$ for $N \rightarrow \infty$,
we obtain $\Sigma_j\frac{1}{N} (1 -\xi_i x_i) x_j\xi_j  \sim (1-\xi_i \bar{x_i}-\sqrt{\tau_J}^{-1}\zeta^n)m$.
$\xi_i \bar{x_i}$ is evaluated as $\xi_i \bar{x_i} = \tanh(\beta \xi_i (J(x^0+z)+\eta)_i)=\tanh(\beta (s_i + \sum_j J_{ij} (\xi_j \epsilon_j+z_j)) $,
where $\epsilon_i$ is deviation of the current state from $\xi_i$ satisfying $x^0_i=\xi_i(1-\epsilon_i)$ and $z_j$ is a random variable sampled from $\mathcal{N}(0,D_n)$.
Further, by mean-field approximation, we derive $\xi_i \bar{x_i} = \int \tanh(\beta u) P(u|s_i) du$, denoted as $<\tanh(\beta u)>_{u|s_i}$ where $P(u|s_i) =\mathcal{N}(s_i,\Sigma^2=R(D_n + 2(1-m))$.
We use $<\epsilon_i>_i=1-m$ and   $<(\xi_i \epsilon_i)^2>_i=2(1-m)$,
assuming $x^0_i$ is around $\pm 1$ due to high value of $\beta (=10)$ and, consequently, $\epsilon_i \in \{0,2\}$.
Thus,  $$\sum_j\frac{1}{N} (1 -\xi_i x_i) x_j\xi_j  \sim (1-<\tanh(\beta u)>_{u|s_i})m + \frac{m}{\sqrt{\tau_J}}\zeta^n$$.

\subsection*{Appendix 3: Dependence of $m$ and $p_c$ on $D_s$ and $D_n$}

We analyze how neural and synaptic noise amplitudes affect the learning performance $m$ and the robustness $p_c$ as described in Eqs. \eqref{eq:m1} and \eqref{eq:p_c}.
First, we evaluate $m$ approximately as the one-step self-consistent iteration $m^{(1)} = \int \langle \tanh(\beta u) \rangle_{u|s} P_{\text{st}}(s|m^{(0)}) ds$, where $m$ is initialized as $m^{(0)} = 1$.
The probability density function is given by $P_{\text{st}}(s|m^{(0)}) = \mathcal{Z}^{-1} \exp(-V(s|m^{(0)})/D')$ as shown in Eq. \eqref{eq:potential}.
In the following, $P_{\text{st}}(s|m^{(0)})$ and $V(s|m^{(0)})$ are referred to simply as $P(s)$ and $V(s)$, respectively.
Considering $\langle V'(s) \rangle = \int_{-\infty}^{\infty} V'(s)P(s)ds = - \int_{-\infty}^{\infty} D'P'(s) ds = 0$ and $V'(s)=-1+\langle\tanh(\beta u)\rangle_{u|s} +\gamma s$, we derive
\begin{equation*}
m^{(1)} = 1 - \gamma \langle s \rangle ,
\end{equation*}
where $\langle s \rangle = \int s P(s) ds$.

Now, we roughly evaluate $\langle s \rangle$ with $m=1$.
The potential $V(s)$ in Eqs. \eqref{eq:stationary_dist} and \eqref{eq:potential} consists of a restoring force ($\gamma s^2/2$) and a pushing force (the remaining terms of $V(s)$).
We approximate $\langle s \rangle$ as the mean of the half-Gaussian part of $P(s)$ for $s^*<s$, shifted by $s^*$:
\begin{eqnarray}
\langle s \rangle \approx s^* + \sqrt{\frac{2D'}{\pi \gamma}}.
\label{eq:<s>}
\end{eqnarray}
because the width of the parabolic part is larger than $s^*$ except in the case $D_s \sim 1$ and $D_n \sim 1$.

Next, we evaluate the dependence of $p_c$ on noise by estimating the rising point $s_r < s^*$ of $P(s)$, defined by $V(s_r) - V(s^*) = D'$, where $p_c \propto s_r$.
By expanding the potential derivative to first order, $V'(s) \approx ks - 1$ with an effective stiffness $k \approx \sqrt{2/\pi}\Sigma^{-1}$, the left side of $P(s)$ ($s<s^*$) is approximated as a half-Gaussian distribution centered at $s^* \approx \sqrt{\pi/2}\Sigma$.
Solving $k(s_r - s^*)^2 / 2 = D'$ for the left root yields:
\begin{eqnarray}
s_r \approx \sqrt{\frac{\pi}{2}}\Sigma - \sqrt{\sqrt{2\pi} D' \Sigma}.
\label{eq:s_r}
\end{eqnarray}

Finally, we roughly evaluate the scaling of $1-m^{(1)}$ and $p_c$ with respect to $D_s$ and $D_n$.
To determine an effective exponent, we evaluate $\partial \ln  \{\langle s \rangle\ \text{or } s_r\} / \partial \ln D_{s,n}$ for $D_{s,n} \sim 1$.
Then, we obtain numerically $\langle s \rangle \sim D_n^{0.38}$  and $\langle s \rangle \sim D_s^{0.63}$  .
Similarly, the scaling of $s_r$ with respect to $D_n$ is approximately $0.4$ and $0.5$ for $D_s =10^{-2}$ and $1$.
For $D_n=0$, $s_r$ is always zero independent of $D_s$, while for $D_n=1$, $s_r \propto D_s^{0.18}$.

\end{document}